\documentclass[prd,superscriptaddress,amsfonts,amssymb,amsmath,showpacs,twocolumn]{revtex4-2}
\usepackage{bm}
\usepackage{amsfonts}
\usepackage{latexsym}
\usepackage[latin1]{inputenc}
\usepackage{graphicx}
\usepackage{amsmath}
\usepackage{palatino}
\usepackage{mathpazo}
\usepackage{textcomp}
\usepackage{float}
\usepackage{booktabs}
\usepackage{dcolumn}
\usepackage{ragged2e}
\usepackage{hyperref}
\hypersetup{colorlinks,citecolor=green}
\hypersetup{colorlinks=true,linkcolor=blue,filecolor=blue,    urlcolor=magenta}
\usepackage{amsmath}
\usepackage{xcolor}
\usepackage[caption=false]{subfig}
\usepackage{commath}
\def\jnl@style{\it}
\def\aaref@jnl#1{{\jnl@style#1}}

\def\aaref@jnl#1{{\jnl@style#1}}

\def\aj{\aaref@jnl{AJ}}                   
\def\apj{\aaref@jnl{ApJ}}                 
\def\apjl{\aaref@jnl{ApJ}}                
\def\apjs{\aaref@jnl{ApJS}}               
\def\apss{\aaref@jnl{Ap\&SS}}             
\def\aap{\aaref@jnl{A\&A}}                
\def\aapr{\aaref@jnl{A\&A~Rev.}}          
\def\aaps{\aaref@jnl{A\&AS}}              
\def\mnras{\aaref@jnl{Mon.~Not.~Roy.~Astron.~Soc.}}             
\def\prd{\aaref@jnl{Phys.~Rev.~D}}        
\def\plb{\aaref@jnl{Phys.~Lett.~B}}        
\def\prc{\aaref@jnl{Phys.~Rev.~C}}  
\def\prl{\aaref@jnl{Phys.~Rev.~Lett.}}    
\def\qjras{\aaref@jnl{QJRAS}}             
\def\skytel{\aaref@jnl{S\&T}}             
\def\ssr{\aaref@jnl{Space~Sci.~Rev.}}     
\def\zap{\aaref@jnl{ZAp}}                 
\def\nat{\aaref@jnl{Nature}}              
\def\aplett{\aaref@jnl{Astrophys.~Lett.}} 
\def\apspr{\aaref@jnl{Astrophys.~Space~Phys.~Res.}} 
\def\physrep{\aaref@jnl{Phys.~Rep.}}      
\def\physscr{\aaref@jnl{Phys.~Scr}}       
\def\commat{\aaref@jnl{Comm.~Math.~Phys.}}              
\def\science{\aaref@jnl{Science}}               
\def\cqg{\aaref@jnl{Classical Quant.~Grav.}}            
\def\jpcs{\aaref@jnl{JPCS}}                                     
\def\ijmpd{\aaref@jnl{Int.~J.~Mod.~Phys.~D}}                    
\def\grg{\aaref@jnl{Gen.~Relat.~Gravit.}}               
\def\rpp{\aaref@jnl{Rep.~Prog.~Phys.}}          
\def\npa{\aaref@jnl{Nucl.~Phys.~A}}        
\def\lrr{\aaref@jnl{Living Rev.~Rel.}}                   
\def\jcap{\aaref@jnl{J.~Cosmology Astropart.~Phys.}}    
\def\rmp{\aaref@jnl{Rev.~Mod.~Phys.}}   
\def\epjc{\aaref@jnl{Eur.~Phys.~J.~C}}

\allowdisplaybreaks[1]

\begin{document}

\color{black}       

\title{The criteria of the anisotropic quark star models in Rastall gravity  }

\author{Takol Tangphati}  
\email{takoltang@gmail.com}
\affiliation{
Theoretical and Computational Physics Group, \\
Theoretical and Computational Science Center (TaCS), Faculty of Science, \\
King Mongkut's University of Technology Thonburi, 126 Prachauthid Rd., Bangkok 10140, Thailand
}

\author{Ayan Banerjee} 
\email{ayanbanerjeemath@gmail.com}
\affiliation{Astrophysics and Cosmology Research Unit, School of Mathematics, Statistics and Computer Science, University of KwaZulu--Natal, Private Bag X54001, Durban 4000, South Africa}

\author{Sudan Hansraj}
\email{hansrajs@ukzn.ac.za}
\affiliation{Astrophysics and Cosmology Research Unit, School of Mathematics, Statistics and Computer Science, University of KwaZulu--Natal, Private Bag X54001, Durban 4000, South Africa}

\author{Anirudh Pradhan}
\email{pradhan.anirudh@gmail.com}
\affiliation{Centre for Cosmology, Astrophysics and Space Science, GLA University, Mathura-281 406, Uttar Pradesh, India}


\date{\today}

\begin{abstract}
Quark stars are terrestrial laboratories to study fundamental physics at ultrahigh densities and temperatures. In this work, we investigate the internal structure and the physical properties of quark stars (QSs) in the Rastall gravity. Rastall gravity is considered a non-conservative theory of gravity, which is an effective gravity theory at high energy density, e.g., relevant to the early universe and dense, compact objects. We derive the hydrostatic equilibrium structure for QSs with the inclusion of anisotropic pressure. More specifically, we find the QS mass-radius relations for the MIT bag model.
We focus on the model depending on the Rastall free parameter $\eta$ and examine the deviations from the General
Relativity (GR) counterparts.

\end{abstract}

 \maketitle

\section{Introduction}

In order to construct realistic models capable of representing physically viable configurations of matter requires solving a complicated system of gravitational field equations. In this regard, the standard theory of gravity proposed by Einstein has generated over 120 exact solutions to the field equations for the well studied isotropic fluid case. What has proved elusive is finding exact solutions with realistic equations of state since the isotropy equation is complicated. In the literature, there is no exact solution known for the simplest linear barotropic case. When models display an equation of state, it is often discovered after an assumption of one of the gravitational potentials is made \cite{delgaty,finch,hansraj} and is usually very complicated. Relaxation of the condition of isotropy carries the elevated prospects of finding exact solutions with the caveat of introducing imperfect fluids. 

In the 1970s, some six decades after the first perfect fluid solutions emerged, the concept of pressure anisotropy was introduced through the observations of Ruderman \cite{ruderman}, and Canuto \cite{canuto}. Bowers and Liang \cite{bowers} reported the first such solutions and thereafter numerous exact solutions emerged on account of the now trivial  field equations \cite{ponce,gokhroo,bondi,herrera,dev1,dev2}. Solution finding is trivial because any metric can solve the anisotropic field equations. Such solutions must still be examined against the conditions for physical acceptability. In most cases, there are violations of these conditions, or an equation of state is not realizable.  The possibility of anisotropy carries the major mathematical advantage of studying exact solutions with equations of state, given that two degrees of freedom are inherent in the field equations. Various conjectures have been made regarding the question of what could give rise to anisotropy. For example, pion condensation in neutron stars could account for unequal pressure stresses \cite{sawyer}. It is also known that Boson stars display pressure anisotropy \cite{gleiser}. Therefore there are solid motivations to investigate anisotropic stellar distributions. 

It is widely believed that general relativity may need augmentation at the scale of the universe and that it is also not well tested in extreme gravity regimes such as near black holes. Additionally, the explanation of the accelerated cosmic expansion evades general relativity unless an appeal to exotic matter,  such as dark energy and dark matter, is made. Rastall \cite{Rastall:1972swe,Rastall:1976uh}   argued that the vanishing divergence of the Einstein tensor does not necessarily imply the vanishing divergence of the energy-momentum tensor. There is room for the covariant divergence of the energy-momentum tensor to vary as the gradient of the Ricci scalar in his proposal. In such a scenario, energy conservation must be sacrificed. However, when studying closed compact objects such as stars, it is not unreasonable for net energy production or loss as such are not genuinely isolated systems. Besides, non-conservative systems have been an active and fertile area of research for a long time. This property is a feature of quadratic Weyl conformally invariant gravity \cite{weyl} and its generalisation by Dirac \cite{dirac1,dirac2}, $f(R, T) $ theory \cite{harko}  and trace-free gravity \cite{ellis1,ellis2,hans-ellis}. The theories mentioned above have emerged due to efforts to correct shortcomings in general relativity.

It has been shown that Rastall gravity does admit the accelerated expansion of the universe.   The basic argument is that the nonconserved elements of the energy-momentum tensor play the roles of dark energy to support cosmic inflation \cite{rawaf,batista,fabris,bronnikov,batista2,moradpour}. Additionally, it has been shown that  stellar models are well behaved in theory \cite{hans-ban1,hans-ban2}.   At the geometric level, Rastall gravity is equivalent to Einstein gravity \cite{visser}, unimodular theory as well as $f(R, T)$ theory; however, the point that the physical consequences are dramatically different has been amplified several times \cite{darabi}.  The one notable negative feature of Rastall gravity is the absence of Lagrangian - various efforts at constructing a suitable Lagrangian have failed thus far, and some wonder if this is even possible. Nevertheless, the many positive features of the theory cannot be ignored. Various aspects of this theory including theoretical and observational once have been reported in recent literature 
\cite{33a,33b,33c,33d}.

To create a configuration for QS in the current investigation that complies with the available observational data on the $(M-R)$ relations, we use the quark matter as our primary input. The paper is formatted as follows: For a static and spherically symmetric line element, we describe the issue and derive the associated field equations using Rastall gravity in Section \ref{sec2}. Section \ref{sec3} is devoted to prescribing a quark matter EoS with proper boundary conditions.
We numerically solve these equations and provide our critical findings in Section \ref{sec4}. We also look at the stability of compact configurations offered by the MIT bag model EoS in this part.
In the conclusion section \ref{sec5}, we summarise the findings. 


\section{Field equations in Rastall gravity model}\label{sec2}

In this section, we introduce fundamental aspects of Rastall gravity theory. As explained above, the possibility that the divergence of the energy-momentum tensor is proportional to the gradient of the Ricci scalar without violating the vanishing divergence of the Einstein tensor also exists. That is Rastall postulated the condition $\nabla_b T^{ab} \propto \nabla R^b$.  Effectively the gravitational field equations may be written in the form \cite{Oliveira:2015lka,Rastall:1972swe,Rastall:1976uh}

\begin{equation}
    G_{\mu \nu} - \gamma g_{\mu\nu} R = 8 \pi G T_{\mu\nu}, 
\label{Rastall_field}
\end{equation}
where $\gamma \equiv \left( \eta - 1 \right)/2$ is the Rastall parameter and $\eta$ is the Rastall free parameter. $G_{\mu \nu}$ and $T_{\mu \nu}$ are the Einstein tensor and the energy-momentum tensor, whereas $g_{\mu \nu}$ is the metric tensor, $R$ is the scalar curvature, and $G$ is Newton's gravitational constant, respectively.

The Rastall field equations from Eq.~(\ref{Rastall_field})  reduce to the standard general relativity (GR) equations of motion when $\eta = 1$. It is also noteworthy that the conservation law of the energy-momentum tensor is no longer valid, and the divergence of the energy-momentum tensor  given by
\begin{eqnarray}
    \nabla^{\mu} T_{\mu \nu} = \frac{1}{2} \left( \frac{\eta - 1}{2\eta - 1} \right) \nabla_{\nu} T \neq 0
    \label{non_conserved}
\end{eqnarray}
is sometimes called the non-conservation of the energy equation. 
We avoid the value $\eta = 1/2$ since the divergence becomes singular in this case. In the following context, we choose units with $G = c = 1$ throughout the paper.

In the current investigation, we consider the anisotropic form of the energy-momentum tensor given by 
\begin{eqnarray}
T_{\mu \nu} = (\rho + p_{\perp})u_\mu u_\nu + p_{\perp}g_{\mu \nu} + (p_{r} - p_{\perp}) \chi_{\mu} \chi_{\nu}, \label{eq3}
\end{eqnarray}
where $\rho$ is the energy density, $p_r$ is the radial pressure, $p_{\perp}$ is the tangential pressure, $u^{\mu}$ is the 4-velocity of the fluid, and $\chi^{\mu}$ is the space-like unit vector in the radial direction. 
One can adjust the field equations of Rastall gravity from Eq.~(\ref{Rastall_field}) into the GR field equation as follows \cite{Oliveira:2015lka},
\begin{eqnarray}
    G_{\mu \nu} = 8 \pi \Tilde{T}_{\mu \nu},
\end{eqnarray}
where $\Tilde{T}_{\mu \nu}$ is the effective energy-momentum tensor,
\begin{eqnarray}
    \Tilde{T}_{\mu \nu} = T_{\mu \nu} - \frac{1}{2} \left( \frac{\eta - 1}{2\eta - 1} \right) g_{\mu \nu} T.
    \label{T_eff}
\end{eqnarray}
This leads to a novel version of the conservation law of the  energy-momentum tensor for Rastall gravity given by
\begin{eqnarray}
    \nabla^{\mu} \Tilde{T}_{\mu \nu} = 0.
\end{eqnarray}

The effective energy density $\Tilde{\rho}$, radial pressure $\Tilde{p}_r$, and tangential pressure $\Tilde{p}_{\perp}$ can be written in terms of the traditional energy density $\rho$, radial pressure $p_r$, and tangential pressure $p_{\perp}$ as follows,
\begin{eqnarray}
    \Tilde{\rho} &=& \frac{1}{2} \left( \frac{3\eta - 1}{2 \eta - 1} \right) \rho + \frac{1}{2} \left( \frac{\eta - 1}{2 \eta - 1} \right) p_r + \left( \frac{\eta - 1}{2 \eta - 1} \right) p_{\perp}, \nonumber\\
    \\
    \Tilde{p}_r &=& \frac{1}{2} \left( \frac{\eta - 1}{2 \eta - 1} \right) \rho + \frac{1}{2}\left( \frac{3\eta-1}{2\eta-1}\right)p_r + \frac{1}{2}\left( \frac{1 - \eta}{1 - 2\eta} \right) p_{\perp}, \nonumber \\
    \\
    \Tilde{p}_{\perp} &=& \frac{1}{2} \left( \frac{\eta - 1}{2\eta - 1} \right) \rho + \frac{1}{2} \left( \frac{1-\eta}{2 \eta -1} \right) p_r + \left( \frac{\eta}{2\eta-1} \right) p_{\perp} \nonumber \\
\end{eqnarray}

We apply the static and spherically symmetric spacetime metric into the Rastall field equations for investigating the properties of the QSs,
\begin{eqnarray}
    ds^2 = -e^{\nu(r)} dt^2 + e^{\lambda(r)} dr^2 + r^2 \left( d \theta^2 + \sin^2 \theta d \phi^2 \right),
    \label{line_element}
\end{eqnarray}
where $\nu(r)$ and $\lambda(r)$ are functions of the radial coordinate, $r$.

One obtains the modified TOV equations by using the modified field equation (\ref{Rastall_field}) and the non-conservation energy-momentum tensor (\ref{non_conserved}),
\begin{eqnarray}
    \frac{d \Tilde{p}}{dr} &=& - \frac{\left( \Tilde{\rho} + \Tilde{p}_r \right) \left( \Tilde{M} + 4 \pi r^3 \Tilde{p}_r \right)}{r^2 \left( 1 - \frac{2 \Tilde{M}}{r} \right)} + \frac{2\left( \Tilde{p}_{\perp} - \Tilde{p}_r \right)}{r}, \label{dpdr} \\
    \frac{d \Tilde{M}}{dr} &=& 4 \pi \Tilde{\rho} r^2,
    \label{dMdr}
\end{eqnarray}
where the corresponding functions in the metric tensor are given by
\begin{eqnarray}
    \frac{d \nu}{dr} &=& - \frac{2}{\Tilde{p} + \Tilde{\rho}} \left(  \frac{d \Tilde{p}}{dr} - \frac{2\left( \Tilde{p}_{\perp} - \Tilde{p}_r \right)}{r} \right), \\
    e^{-\lambda} &\equiv& 1 - \frac{2 \Tilde{M}}{r}.
\end{eqnarray}
where $\Tilde{M}$ denotes the  effective mass (not physical mass). To analyze the physical quantities,  the following mass definition \cite{Velten:2016bdk, Maulana:2019}
\begin{eqnarray}
    M \equiv \int 4\pi r^2 \rho(r) dr
\end{eqnarray}
is utilised. 



\section{Boundary constraints and the star's exterior vacuum area } \label{sec3}

\subsection{Anisotropic MIT bag model equation of state} 

Usually, quarks cannot be stable as free particles because of 'absolute confinement' unless the confinement of two or three quarks is to synthesize a meson or a hadron, respectively. According to Quantum chromodynamics (QCD), a phase transition between hadrons and the deconfined quarks and gluons, also known as quark-gluon plasma (QGP), is studied at high energy density and temperature. The investigation in QCD provides an exciting phase where matter fields are strongly interacting. This leads to the possibility for the quark matter state to exist at the center of the massive neutron stars (NSs). Current studies on the strange quark matter (SQM) as the candidate for the core of the compact stars in \cite{Witten:1984rs,Bodmer:1971we,Itoh1970}. SQM predicts the quark matter core consists of almost the same amount of $u$, $d$, and $s$ quarks where the amount of $s$ quarks is fewer since it is relatively higher mass. Because the energy per baryon of SQM is lower than that of the most stable atomic nuclei, like $^{56}$Fe and $^{62}$Ni, SQM is implied to be one of the most stable quantum states of the hadronic matter \cite{Bodmer:1971we,Itoh1970}.

In this research, we apply the simplest and most  well-known model of SQM called the MIT bag model. This model utilizes the confinement properties in QCD to describe the quark-based hadron properties. The baryon in the bag model is made up of 3 non-interacting quarks with the required energy for creating the bag $B$ \cite{Flores:2017kte}. The EoS of the model can be written in terms of energy density $\rho$, and the radial pressure $p_r$ as follows \cite{Alford:2004pf}
\begin{align} \label{Prad1}
  p_r =&\ \dfrac{1}{3}\left(\rho - 4B\right).
\end{align}
The value of the bag constant $B$, in general,  lies within the range of  $57 \leq B \leq 92$ MeV/fm$^3$ \cite{Burgio:2018mcr, Blaschke:2018mqw}. In recent times, many theoretical works have investigated  
the possible formation of QS depending on the EoS (\ref{Prad1}) (see, e.g., Refs. \cite{Becerra-Vergara:2019uzm, Banerjee:2020dad, Panotopoulos:2021sbf}).

For the anisotropic quark matter, the EoS employed in this work was first proposed by Horvat \textit{et al} \cite{Horvat:2010xf}.
Recently, QSs in $R^{1+\epsilon}$ gravity  with pressure
anisotropy were obtained in Ref. \cite{Pretel:2022plg}. Following Refs. \cite{Horvat:2010xf}, we introduce the anisotropy parameter
\begin{eqnarray}
    \sigma &\equiv& p_{\perp} - p_r =  \beta p_r \mu,
\end{eqnarray}
with $\sigma =0$ corresponding to isotropic matter.  The tangential pressure $p_{\perp}$ is given by
\begin{eqnarray}
    p_{\perp} &=& p_r \left( 1 + \beta \left[ 1 - e^{-\lambda} \right] \right),
\end{eqnarray}
where $\mu \equiv 2m(r)/r$ and $\beta$ represents the measure of the anisotropicity which can be positive or negative values \cite{Horvat:2010xf,Doneva:2012rd,Silva:2014fca,Yagi:2015hda,Pretel:2020xuo,Rahmansyah:2020gar,Rahmansyah:2021gzt,Folomeev:2018ioy}. This assumption of the anisotropic quark matter is based on the relativistic limit since the difference between the radial and tangential pressures will deplete in the non-relativistic limit ($\mu \rightarrow 0$). Owing to the quark matter state, the high density at the core of the QSs supports the anisotropicity. Moreover, at the center of the QSs, the stellar fluid becomes isotropic since $\mu \sim r^2$ at $r \rightarrow 0$.


\subsection{The boundary conditions of the quark star}

To unravel the internal structure of QSs in Rastall gravity, we numerically integrate the modified TOV equations (\ref{dpdr}) and (\ref{dMdr}) along with the EoS (\ref{Prad1}). 
At the center of a QS, we determine the boundary conditions as follows,
\begin{align}
   \rho(0) &= \rho_c,  ~~~~    \text{and} ~~~~  m(0) = 0 ,
   \label{BC_Inner}
\end{align}
where $\rho_c$ is the central energy density of QS.
Then we integrate outward until the pressure disappears where the surface of the star is identified when 
\begin{eqnarray}
    p_r(r_{\rm s}) = p_{\perp}(r_{\rm s}) = 0,
    \label{BC_Outer}
\end{eqnarray}
where $r_{\rm s}$ is the radius of the star.

\section{Numerical results and discussion}\label{sec4}

\begin{figure}
    \centering
    \includegraphics[width = 8 cm]{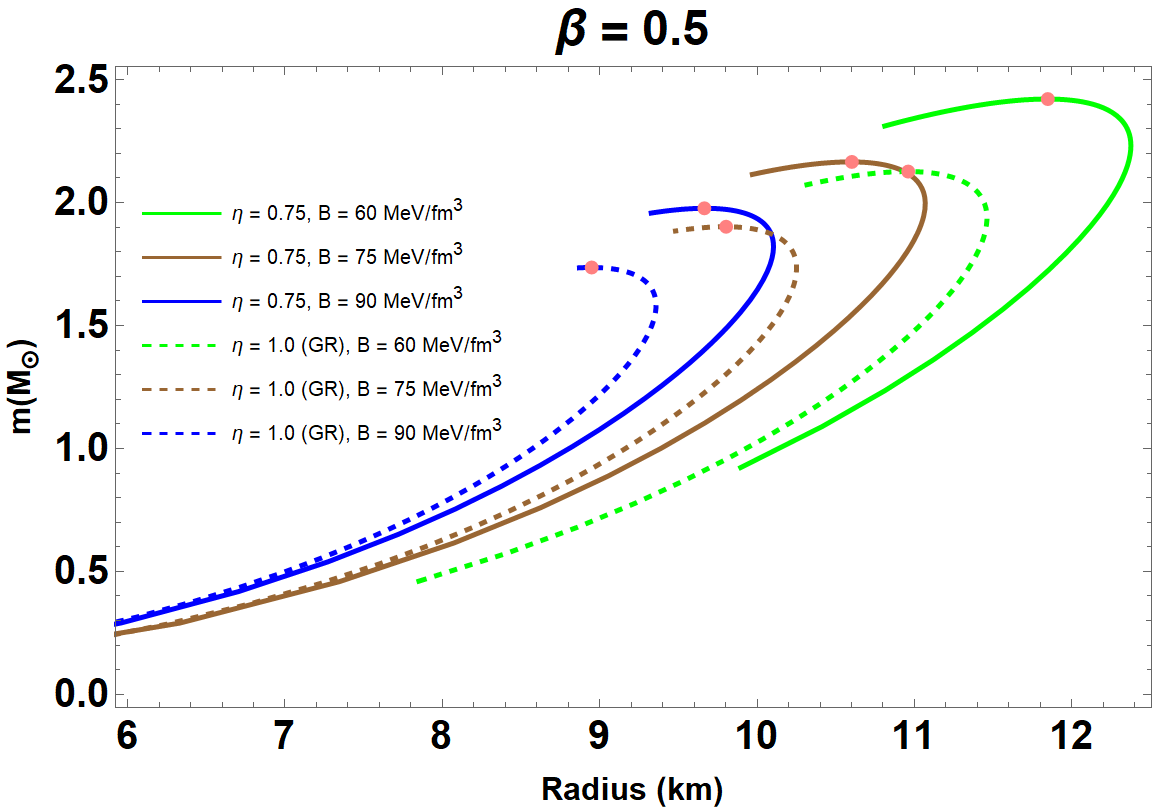}
    \includegraphics[width = 8 cm]{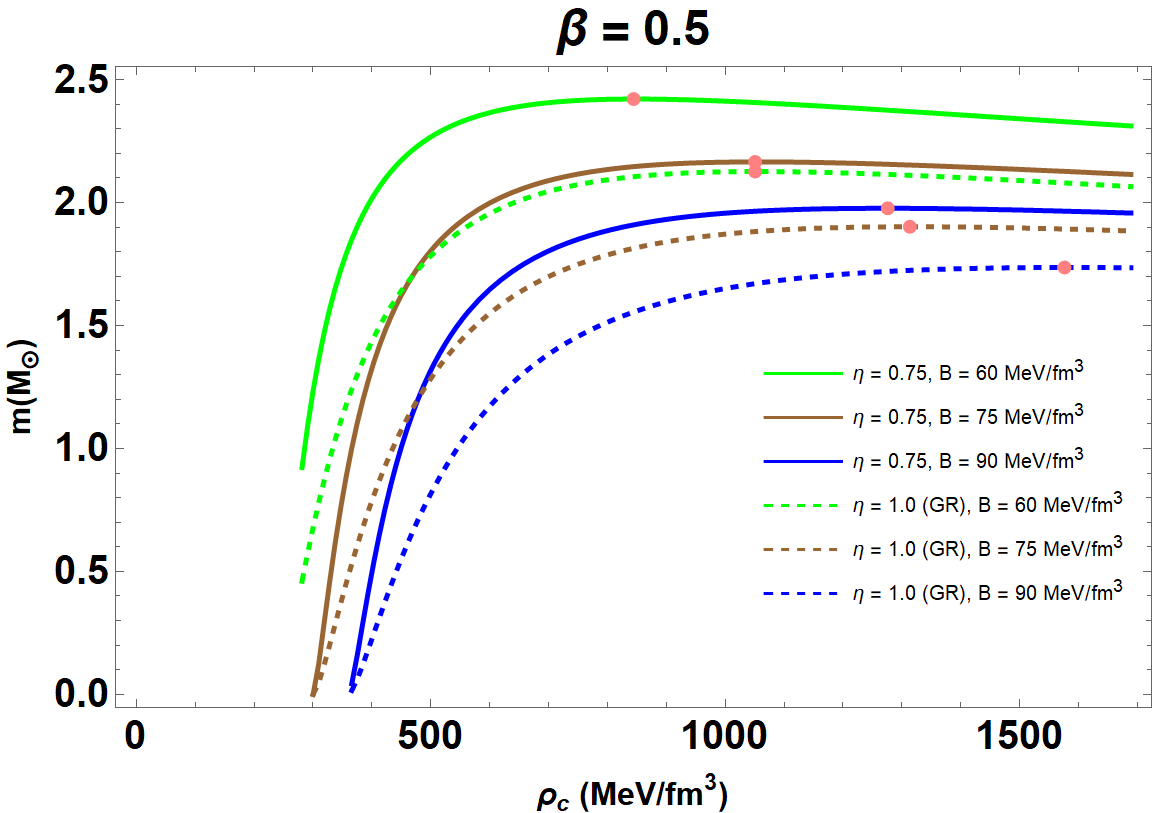}
    \caption{Mass-radius (upper panel) and mass-central energy density (lower panel) relations have been plotted by solving the structure equations (\ref{dpdr}) and (\ref{dMdr}) along with the EoS (\ref{Prad1}). For the QS model, we vary $B \in [60,90]$ MeV/fm$^3$ with the other sets of parameters remaining fixed $\beta = 0.5$ and $\eta = 0.75$. Here, the colored curves present the anisotropic QS in Rastall gravity while the dashed colored curves present the results for GR.}
    \label{MR_ME_varyB}
\end{figure}

\begin{figure}
    \centering
    \includegraphics[width = 8 cm]{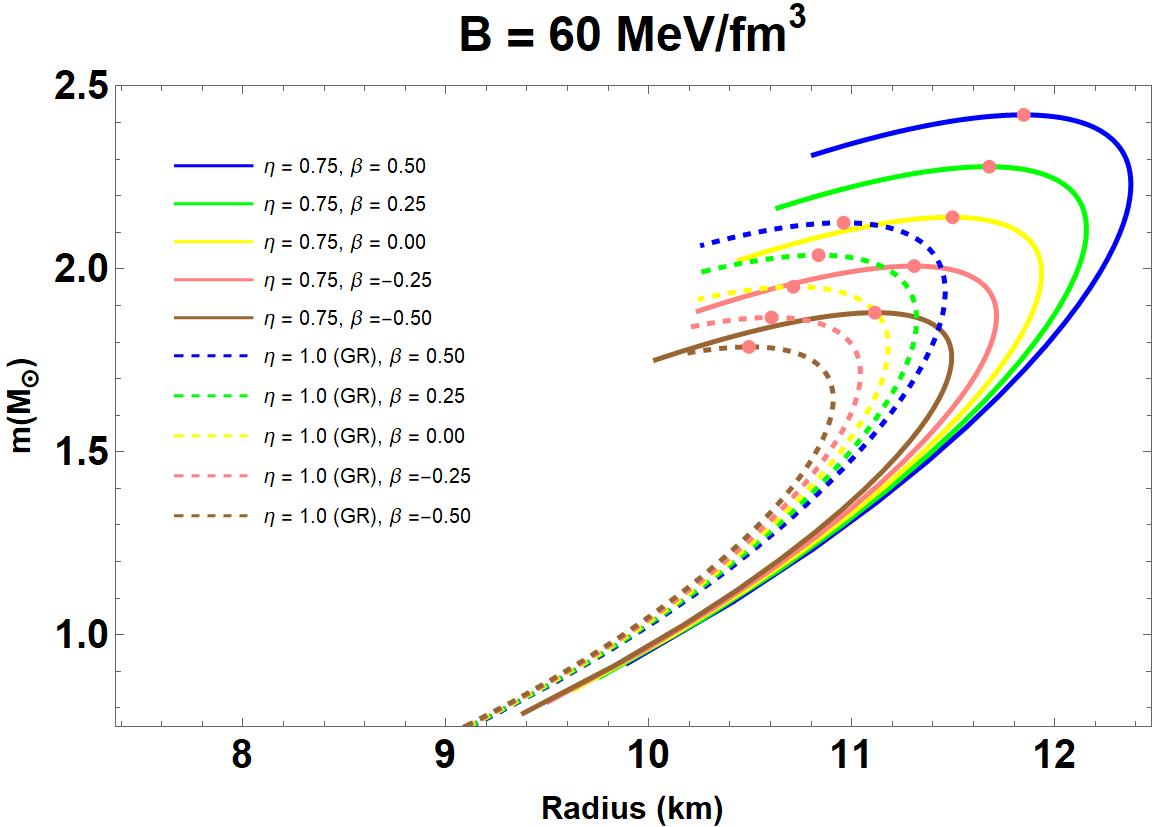}
    \includegraphics[width = 8 cm]{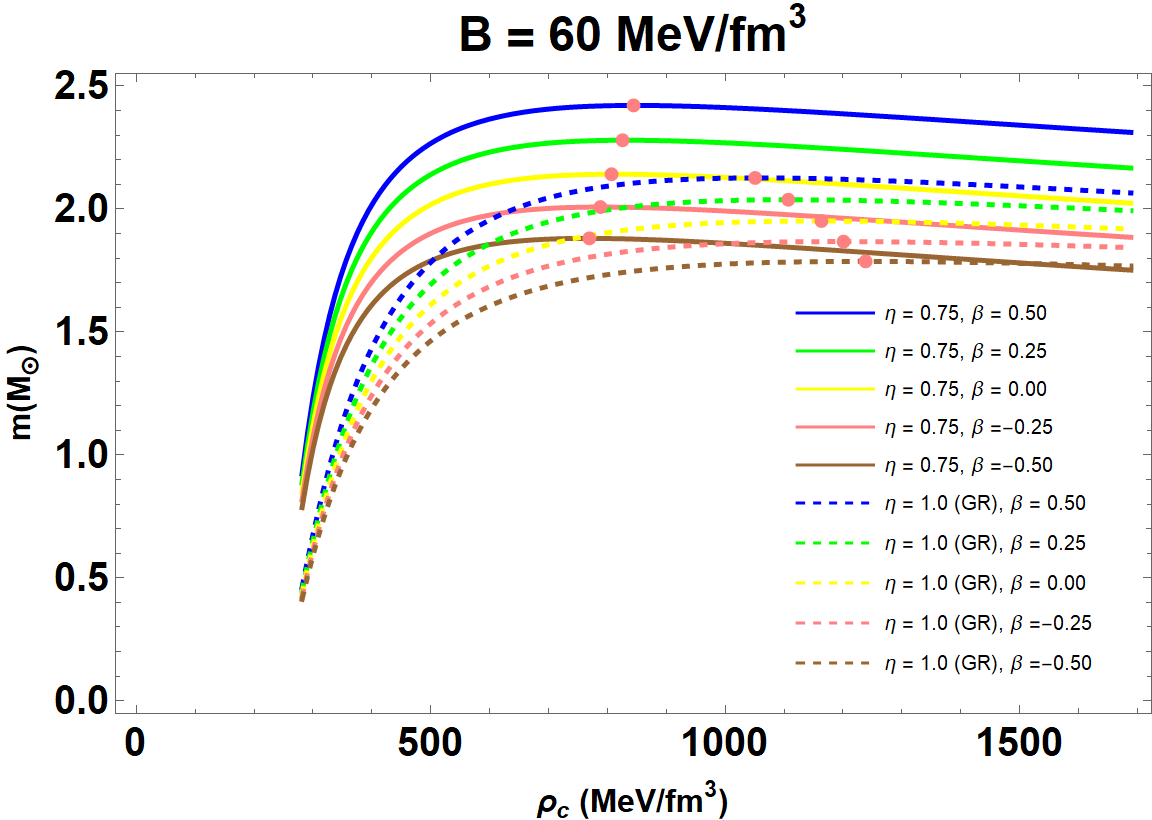}
    \caption{Mass-radius (upper panel) and mass-central energy density (lower panel) relations have been plotted. Here, we vary the anisotropicity parameter $\beta \in [-0.50, 0.50]$ with the other set of parameters remaining fixed $B = 60$ MeV/fm$^3$ and $\eta = 0.75$. The colored curves present the anisotropic quark star in Rastall gravity while the dashed colored curves present the results for GR. }
    \label{MR_ME_varyBeta}
\end{figure}

\begin{figure}
    \centering
    \includegraphics[width = 8 cm]{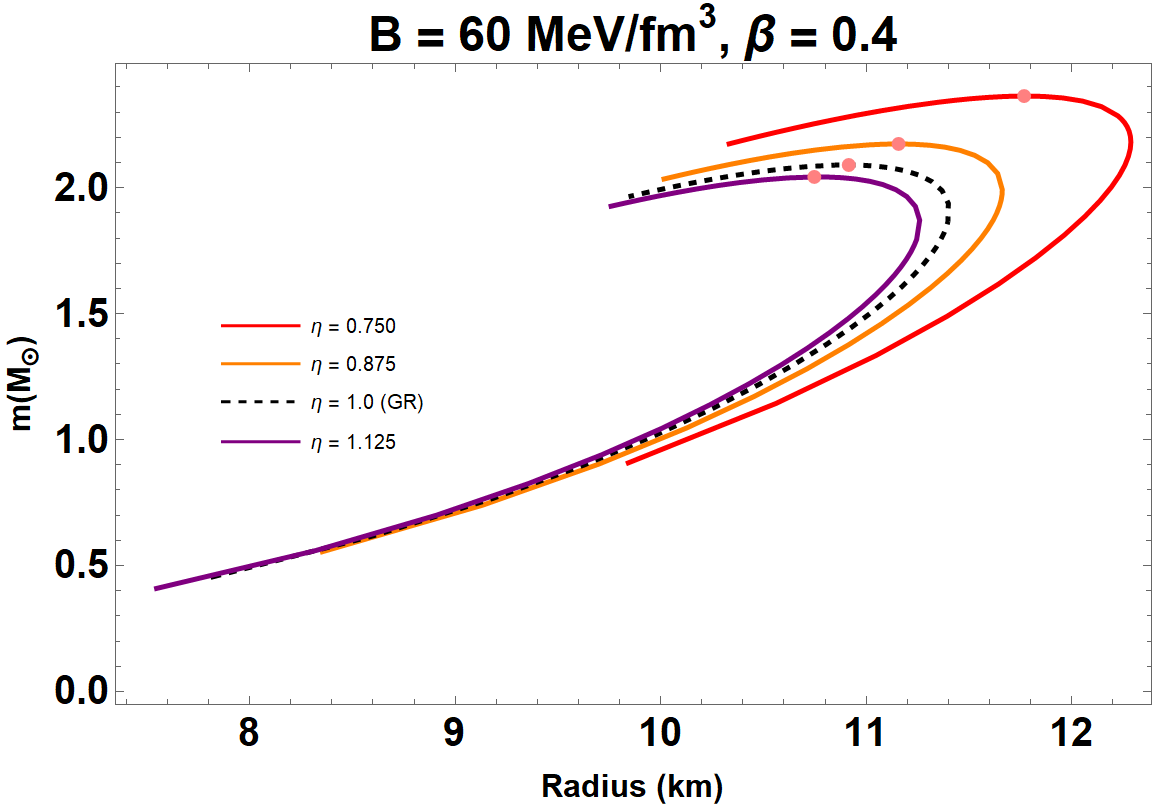}
    \includegraphics[width = 8 cm]{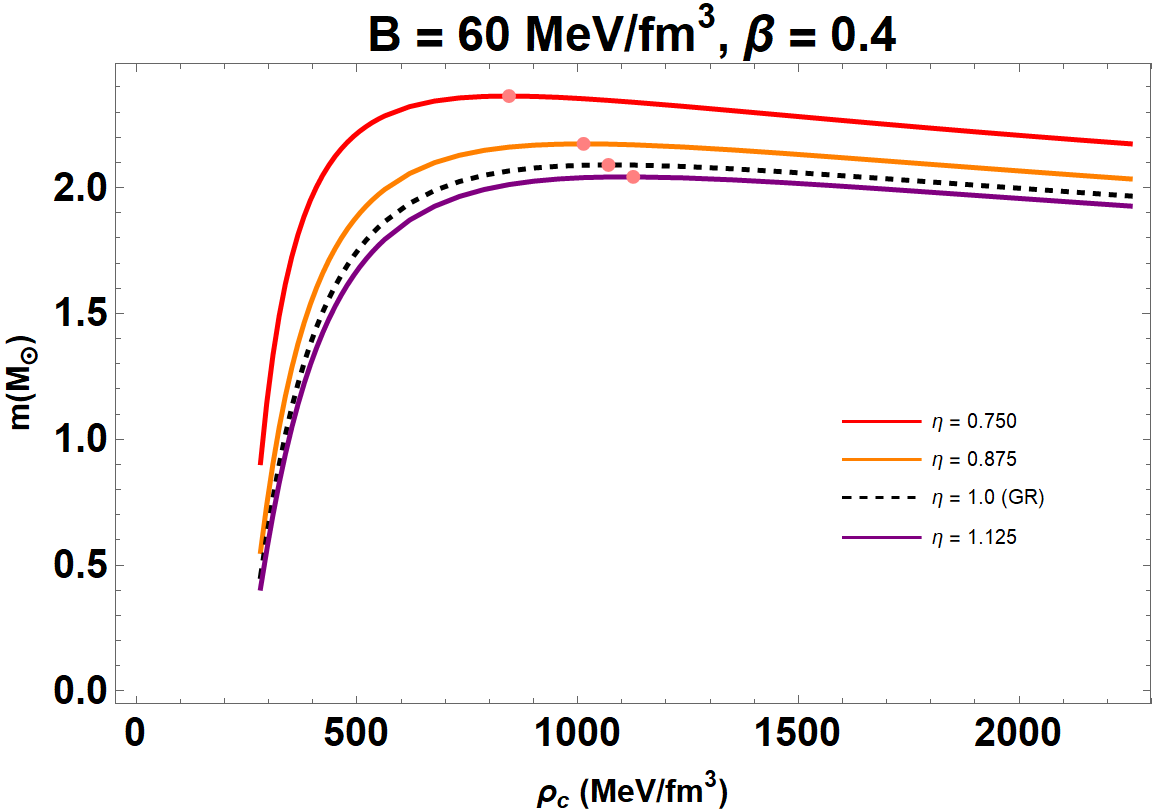}
    \caption{Mass-radius (upper panel) and mass-central energy density (lower panel) relations have been plotted.  The colored curves present the anisotropic quark star in Rastall gravity for $\eta \in [0.75, 1.125]$  with the other sets of parameters remaining fixed for $\beta = 0.4$ and $B = 60$ MeV/fm$^3$. The black dashed curve presents the results for GR ($\eta =1$).  }
    \label{MR_ME_varyEta}
\end{figure}

In the following, we present our results for strange QSs using an EoS presented in the previous subsection. To achieve the properties of  QSs, we apply the numerical method via Mathematica code to integrate Eqs.~(\ref{dpdr})-(\ref{dMdr}) with the proper boundary conditions given in Eqs. (\ref{BC_Inner})-(\ref{BC_Outer}). In this section, we present the numerical results with three groups of the parameter sets, separated by the parameter variation: (i) the Bag constant $B \in [60,90]$ MeV/fm$^3$ as illustrated in Fig.~\ref{MR_ME_varyB}, (ii) the level of the anisotropicity $\beta \in [-0.50, 0.50]$ as illustrated in Fig.~\ref{MR_ME_varyBeta} and (iii) the Rastall free parameter $\eta \in [0.750, 1.125]$ as illustrated in Fig.~\ref{MR_ME_varyEta} separately. These numerical results are presented in the profiles of $M-R$, $M-\rho_c$, and compactness $(M-M/R)$. We also simulate the general relativity (GR) results to compare with the novel numerical findings generated by the Rastall gravity theory. The properties of the QSs, for instance, mass, pressure, central energy density, and compactness, are enlisted in Tables \ref{tableVaryB}, \ref{tableVaryBeta}, and \ref{tableVaryEta}, respectively. It is noteworthy that the mass unit is of solar mass $M_{\odot}$, the radius unit is  ${\rm km}$, and the unit of pressure and energy density is of MeV/fm$^3$.

\subsection{The mass-radius relations}

\begin{table}[h]
    \centering
    \caption{Structure properties of QSs in Rastall gravity with parameters mentioned in Fig.~\ref{MR_ME_varyB}. The compactness ($M/R$) of QS is a dimensionless quantity. }
    \begin{ruledtabular}
    \begin{tabular}{ccccc}
    $B$  & $M$ & $R$ & $\rho_c$ & $M/R$\\
    MeV/fm$^3$ & $M_{\odot}$ &   km & MeV/fm$^3$ & \\
    \hline
        60 &  2.421 & 11.85 & 844 & 0.303 \\
        75 &  2.165 & 10.60 & 1050 & 0.303 \\
        90 &  1.977 & 9.66 & 1275 & 0.303 
    \end{tabular}
    \end{ruledtabular}
    \label{tableVaryB}
\end{table}

\begin{table}[h]
    \caption{Structure properties of QSs in Rastall gravity with parameters mentioned in Fig. \ref{MR_ME_varyBeta}.  The compactness ($M/R$) of QS is a dimensionless quantity. }
    \begin{ruledtabular}
    \begin{tabular}{ccccc}
    $\beta$  & $M$ & $R$ & $\rho_c$ & $M/R$\\
     & $M_{\odot}$ &   km & MeV/fm$^3$ & \\
    \hline
        0.50 & 2.421 & 11.85 & 844 & 0.303 \\
        0.25 & 2.280 & 11.68 & 825 & 0.289 \\
        0.00 & 2.141 & 11.50 & 806 & 0.276 \\
        -0.25 & 2.008 & 11.31 & 788 & 0.263 \\
        -0.50 & 1.881 & 11.11 & 769 & 0.251
    \end{tabular}
    \end{ruledtabular}
    \label{tableVaryBeta}
\end{table}

\begin{table}[h]
    \caption{Structure properties of QSs in Rastall gravity with parameters mentioned in Fig. \ref{MR_ME_varyEta}. The compactness ($M/R$) of QS is a dimensionless quantity. }
    \begin{ruledtabular}
    \begin{tabular}{ccccc}
    $\eta$  & $M$ & $R$ & $\rho_c$ & $M/R$\\
     & $M_{\odot}$ &   km & MeV/fm$^3$ & \\
    \hline
        0.750 &  2.364  & 11.77 & 844 & 0.298 \\
        0.875 &  2.174  & 11.16 & 1013 & 0.289 \\
        1.000 &  2.091  & 10.91 & 1069 & 0.284 \\
        1.125 &  2.043  & 10.75 & 1125 & 0.282 
    \end{tabular}
    \end{ruledtabular}
    \label{tableVaryEta}
\end{table}

In this section, we present the mass-radius $(M-R)$ relation and the mass-central energy density ($M-\rho_c$) relation in Figs. \ref{MR_ME_varyB} to \ref{MR_ME_varyEta} with the variation of $B$, $\beta$, and $\eta$, respectively. For different values of
central energy density, we plot the numerical profiles of $M-R$ and $M-\rho_c$.

For case (i), the numerical outputs have been shown in Fig.~\ref{MR_ME_varyB}, where the color solid
lines represent the modified TOV equations for Rastall gravity, and the color dashed lines represent the standard GR solutions.
We considered $\beta = 0.5$ and $\eta = 0.75$ as a fixed parameter and vary the Bag constant $B \in [60,90]$ MeV/fm$^3$.
Table \ref{tableVaryB} summarizes the main aspects of the model used in this work. In the upper panel of Fig.~\ref{MR_ME_varyB} we display the $M-R$ relations and observe that the maximum mass and its corresponding radius of QSs increase as the bag constant decreases. For an example, the highest maximum mass
is above 2.421 $M_{\odot}$ with its radius 11.85 km at $B = 60$ MeV/fm$^3$, $\beta = 0.5$ and $\eta = 0.75$ for Rastall gravity,
whereas the maximum mass corresponding to GR ($\eta = 1$) is about 2.126 $M_{\odot}$ with its radius 10.96 for the same parameter set. As we anticipated, the maximum mass and the radius of the QSs in the anisotropic Rastall gravity are larger than the GR counterpart. The $M-\rho_c$ profile is presented in the lower panel of Fig. \ref{MR_ME_varyB}.  From Table \ref{tableVaryB},
 one can see that the trend of the central energy density of the maximum mass increases as the bag constant increases. In particular, the highest value of maximum mass is around $M = 2.421 M_{\odot}$ (for $B = 60$ MeV/fm$^3$) when
its energy density reaches the lowest value at $\rho_c = 844$ MeV/fm$^3$. On the other hand, the lowest value of maximum mass is around $M = 1.977  M_{\odot}$ (for $B = 90$ MeV/fm$^3$) when its energy density reaches the highest value at $\rho_c = 1275$ MeV/fm$^3$.

For case (ii), we show the numerical results for the variation of $\beta \in [-0.50, 0.50]$ and present the result in  Fig.~\ref{MR_ME_varyBeta}. The solid curves represent the Rastall gravity, whereas the dashed curves represent the GR solution.  The upper panel of the Fig.~\ref{MR_ME_varyBeta}
shows our results for the $M-R$ profiles where the maximum mass and radius increase as $\beta$ increases. Table~\ref{tableVaryBeta} shows the highest maximum mass of QSs is about 2.421 $M_{\odot}$ with the radius 11.85 km when $\beta = 0.50$ for Rastall gravity, while the maximum mass corresponding to GR is about 2.12 $M_{\odot}$ with the radius 10.95 km for $\beta = 0.50$. For both gravity theories, we find that the maximum mass and its radius increase as the value of the level of anisotropictiy or $\beta$ increases. Moreover, the QSs are more massive in Rastall gravity than the standard GR. In the lower panel of Fig.~\ref{MR_ME_varyBeta}, we present the profiles of the $M-\rho_c$ relation. From the plot, we see that the central energy density of the maximum mass increases as $\beta$ increases; however, the trend is reversed for the GR counterpart. For instance, in the case of Rastall gravity,  the highest maximum mass of QS has been obtained for $\rho_c = 844$ MeV/fm$^3$, while the lowest maximum mass has been obtained for $\rho_c = 769$ MeV/fm$^3$.
For the GR case, we observe that the central energy density $\rho_c = 1,050$ MeV/fm$^3$ is used to reach the highest maximum mass value of QS, while the lowest maximum mass has been obtained for $\rho_c = 1,200$ MeV/fm$^3$.

For the case (iii), we depict the $M-R$ profiles for different values of $\eta \in [0.750, 1.125]$ in Fig.~\ref{MR_ME_varyEta} and enlist the data in Table~\ref{tableVaryEta}. Note that for $\eta = 1.000$, the Rastall gravity reduces to the standard GR solution, and these results are illustrated in black dashed curves for both panels in Fig.~\ref{MR_ME_varyEta}. Here, it is useful to note that we avoid $\eta = 0.500$, because the Eq.~(\ref{non_conserved}) diverges. In the top panel of Fig.~\ref{MR_ME_varyEta}, we see that the maximum mass and its respective radius increase monotonically with decreasing $\eta$. We see that the maximum mass reaches above 2.364 $M_{\odot}$ with its radius 11.77 km for $\eta = 0.750$. In GR ($\eta = 1.000$), the maximum gravitational mass of QS has been obtained at merely 2.091 $M_{\odot}$ with a radius of 10.91 km. In the lower panel of Fig.~\ref{MR_ME_varyEta}, we plot $M-\rho_c$. One should notice that the trend of  $M-\rho_c$ curves is similar to the  Fig. \ref{MR_ME_varyB} i.e. when we increase the central energy density, the maximum mass decreases, see  Table~\ref{tableVaryEta}.

\subsection{The stability criterion and the compactness} 

\begin{figure}
    \centering
    \includegraphics[width = 7.5 cm]{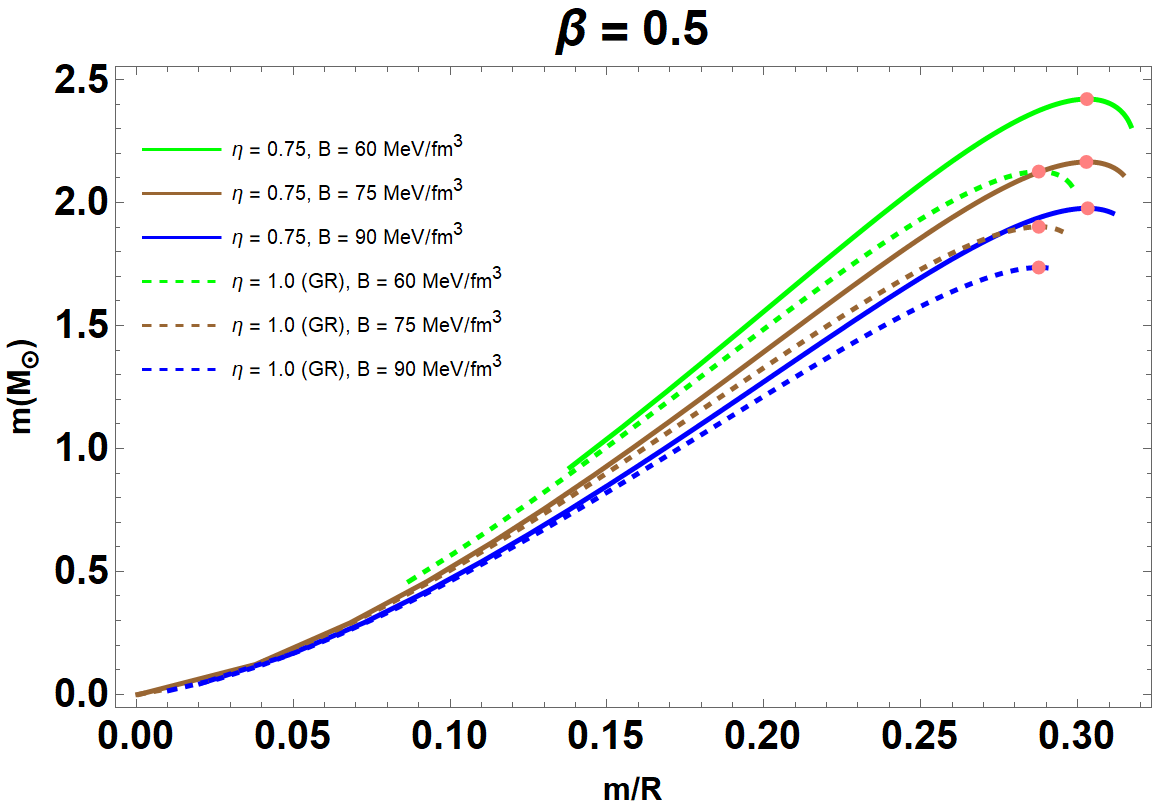}
    \includegraphics[width = 7.5 cm]{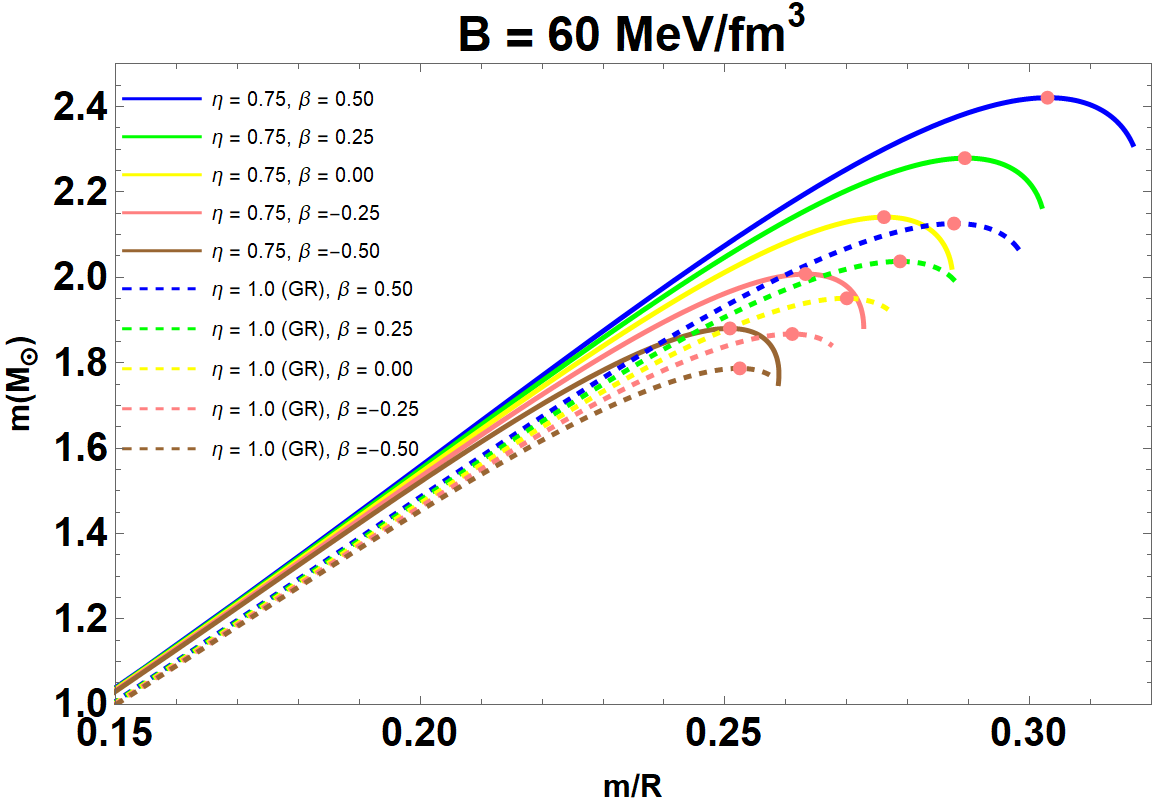}
    \includegraphics[width = 7.5 cm]{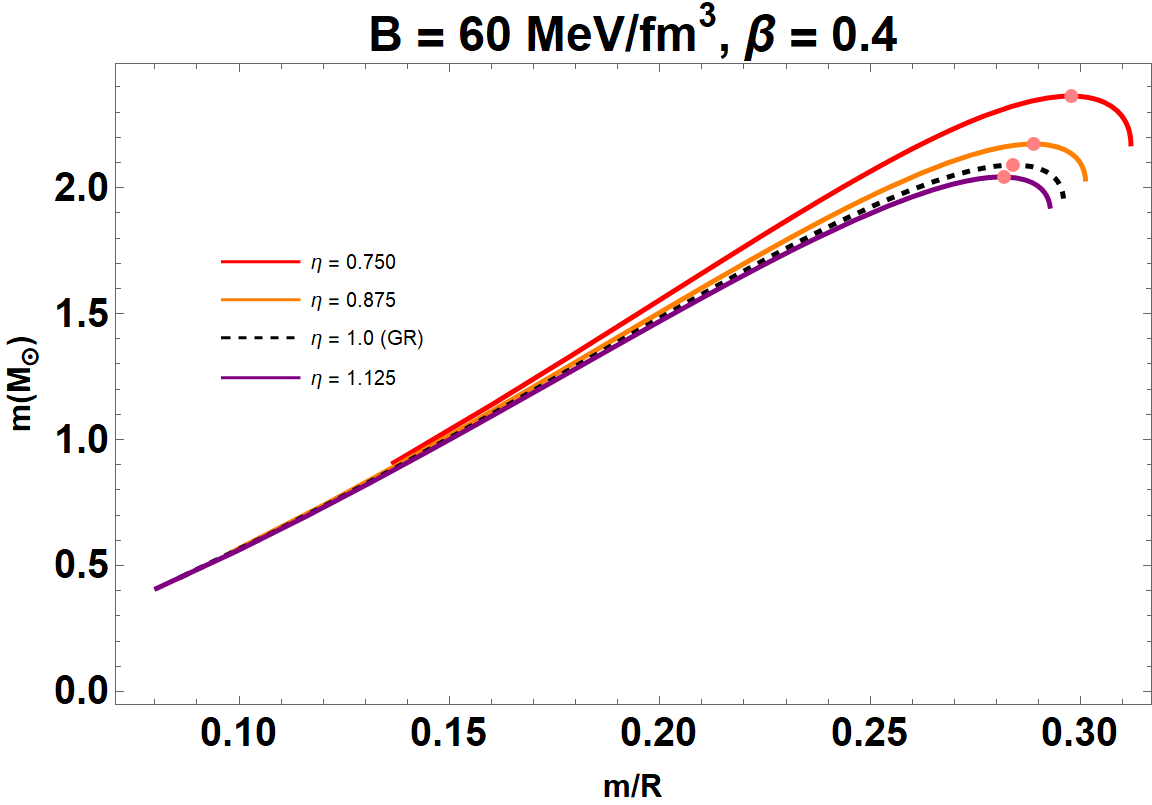}
    \caption{Stellar compactness $M/R$ has been shown against the maximum mass $M$ in solar unit.  The used set of parameters are same as of Fig.~\ref{MR_ME_varyB}, \ref{MR_ME_varyBeta} and \ref{MR_ME_varyEta}, respectively.}
    \label{compactness}
\end{figure}

We now study the stability of QSs in Rastall gravity of the equilibrium configurations. To explore this, we consider a relationship between the total mass $M$ and its central energy density $\rho_c$, known as \textit{static stability criterion} \cite{harrison,ZN}, which is defined by 
\begin{eqnarray}\label{stability}
&&\frac{d M}{d \rho_c} < 0 ~~~ \rightarrow \text{unstable configuration} \\
&&\frac{d M}{d \rho_c} > 0 ~~~ \rightarrow \text{stable configuration},\label{stability1}
\end{eqnarray} 
to be satisfied by all stellar configurations. Since the above conditions are necessary but not sufficient. It is known that the point $(M_{\text{max}}, R_{M_{\text{max}}})$ in the QS $M-R$ relations is a boundary that separate the stable configuration region from the unstable one. As a first step in our analysis, we plot $ M-\rho_c$ curves in Figs.~\ref{MR_ME_varyB}-\ref{MR_ME_varyEta} (in the lower panel).  The $ M-\rho_c$ profiles are in agreement with the $M-R$ profiles, which grow with the increasing values of mass and are stable up to the point where $dM/d\rho_c = 0$. The circle on each $ M-\rho_c$ curve
corresponds to the maximum-mass stellar configuration.

To further explore the structure of QSs, we investigate the compactness ($M/R$) of the specific EoS. Thus the search for the maximum compactness is carried out for the variation of  the MIT bag constant $B \in [60,90]$ MeV/fm$^3$, $\beta \in [-0.50, 0.50]$ and $\eta \in [0.750, 1.125]$ in  Fig.~\ref{compactness}, respectively.  In each figure, the dashed curves represent the GR solution.  In the top panel of Fig. \ref{compactness}, we see that there are no changes in the maximum compactness for the variation of $B$ and the ration of $M/R$
for the Rastall gravity is 0.303, see Table~\ref{tableVaryB}.
In the middle panel of Fig.~\ref{compactness}, we have shown the profile for $M/R$ with the variation $\beta \in [-0.50, 0.50]$.
The maximum compactness increases with the increase of $\beta$ for both gravity theories, i.e., for GR as well as
Rastall gravity. For the same set of parameters, we see the influence of Rastall gravity on the $M/R$ ratio compared with GR. For example, the maximum compactness for Rastall gravity reaches 0.303 (when $\eta = 0.75$) while for GR is only 0.288 (when $\eta = 1.00$), see Table \ref{tableVaryBeta}. In the bottom panel of Fig.~\ref{compactness}, we present the $M/R$ ratio for the variation of $\eta \in [0.750, 1.125]$. It is clear that by increasing $\eta$, the $M/R$ ratio decreases, and the maximum compactness for
Rastall gravity is 0.298 while for GR is just 0.284, see Table \ref{tableVaryEta}.

\subsection{Adiabatic indices}

\begin{figure}[h]
    \centering
    \includegraphics[width = 7.5 cm]{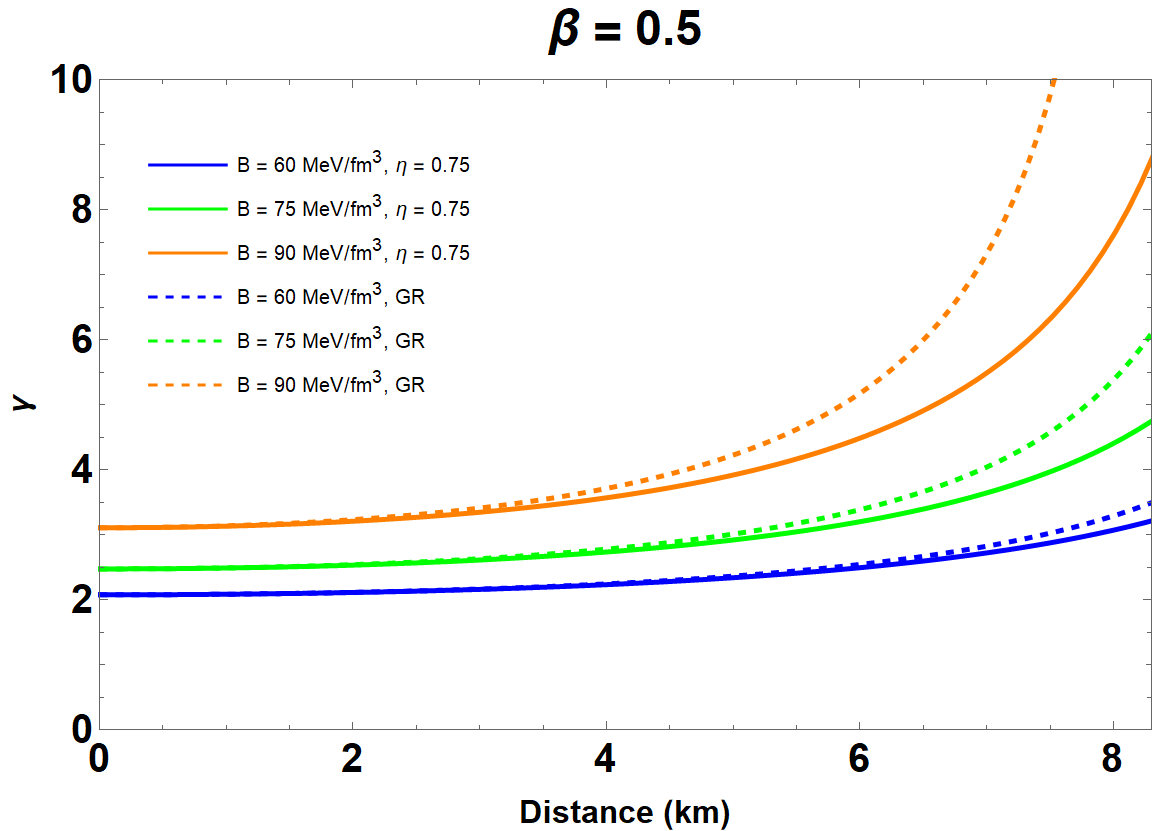}
    \includegraphics[width = 7.5 cm]{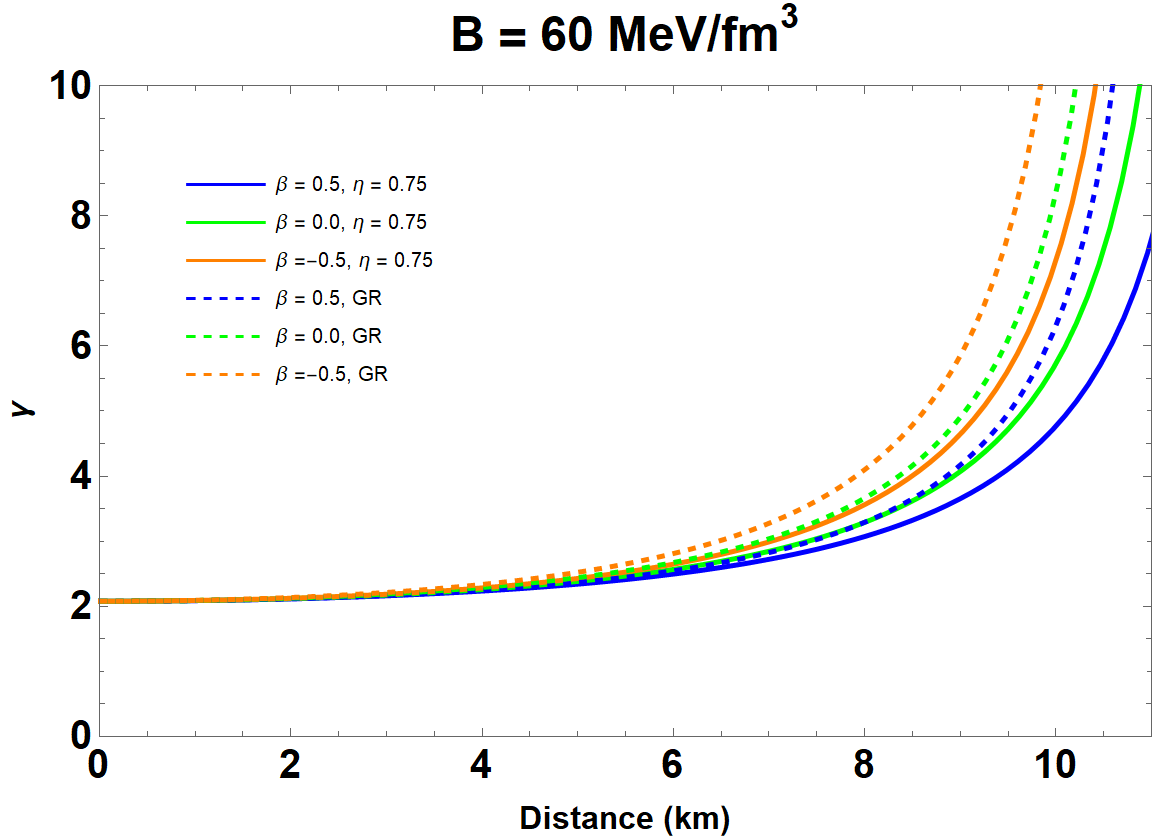}
    \includegraphics[width = 7.5 cm]{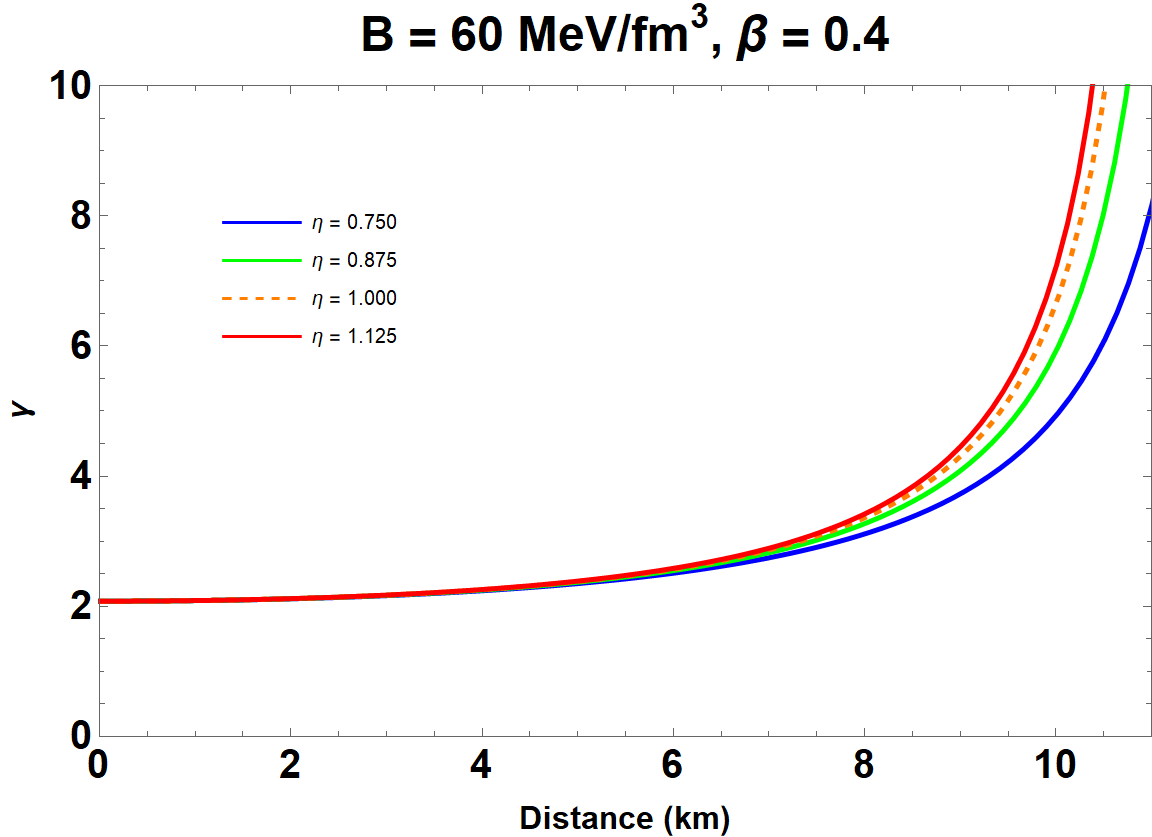}
    \caption{The adiabatic index $\gamma$ has been plotted as a function $r$. The used set of parameters are the same as of Fig.~\ref{MR_ME_varyB}, \ref{MR_ME_varyBeta} and \ref{MR_ME_varyEta}, respectively.}
    \label{fig_gamma}
\end{figure}

In the spirit of dynamical stability based on the variational method, Chandrasekhar \cite{Chandrasekhar}
developed the theory of radial oscillations of relativistic stars. Assuming the adiabatic index  
$\gamma$ of the perturbation which is defined by 
\begin{eqnarray}\label{adi}
    \gamma \equiv\left(1+\frac{\rho}{P}\right)\left(\frac{dP}{d\rho}\right)_S,
\end{eqnarray} 
where  $dP/ d\rho$  is the sound speed and the subscript $S$ indicates the derivation at constant entropy. Here
$\gamma$ is the dimensionless quantity.

The value of $\gamma$ depends on the ratio of $\rho/P$ at the core of the star. If $\gamma > 4/3$ ($\gamma > 4/3$), the star is dynamically stable (unstable) \cite{Glass}. Here,
$\gamma_{\text{cr}} = 4/3$ represents the condition for critical adiabatic index i.e., the critical point of transition from stable to unstable configuration. In Fig. \ref{fig_gamma}, for instance, the adiabatic index $\gamma$ has been plotted as a function $r$ in GR ($\eta = 1) $and in Rastall gravity. The figures clearly indicate that the profile of $\gamma$ is greater than the critical adiabatic index i.e., $\gamma > \gamma_{\text{cr}}$. We thus conclude our model is stable  against the radial adiabatic infinitesimal perturbations.

\section{ concluding remarks }\label{sec5}

We have investigated the effect of pressure anisotropy on compact objects in the Rastall gravity. An important feature of the theory is the non-conservation of the energy-momentum tensor proportionally to the spacetime curvature. This theory can explain the gravitational phenomena at the scale of the Solar System and fit recent cosmological observations. In this text, we consider 
static and spherically symmetric spacetime and assume that the
matter in the stellar interior can be described by the MIT bag model. Finally, we derived the modified TOV equations for a compact star and studied the equilibrium structure in the presence of pressure anisotropy.

Having the EoS for quark matter, we have analyzed the physical properties of dense, compact objects. Depending on a wide range of the parameters ( $B$, $\beta$ and $\eta$), we studied mass-radius
relation ($M-R$), mass-central energy density ($M-\rho_c$) and compactness ($M/R$) for QSs. All the cases we consider have been analyzed effectively for the Rastall gravity model of the star and compared to its GR counterpart. Our analysis shows
that,  by decreasing $\eta $, the maximum mass of QS increases to more than 2 $M_{\odot}$.  For example, when  $\eta = 0.750$, the QS is around 2.364 $M_{\odot}$ for Rastall gravity while for GR it is about 2.091 $M_{\odot}$, See Table \ref{tableVaryEta}. Further, we examine
the effects of the bag constant $B$ and the anisotropy parameter $\beta$  on the QS mass-radius relation. In all of these cases, we have seen that the trend is almost similar to the previous case, which gives rise to more compact stars than in GR ( see Tables~\ref{tableVaryB} and \ref{tableVaryBeta}). Finally,
our analysis confirms that the proposed model is dynamically stable based
on the variational method. We concluded that it is possible to obtain supermassive compact stars in this gravity theory and give a realistic description of $M-R$ relation satisfying the observational
constraints.

\begin{acknowledgments}
 T. Tangphati was supported by King Mongkut's University of Technology Thonburi's Post-doctoral Fellowship.  A. Pradhan expresses gratitude to the IUCCA, Pune, India, for offering facilities under associateship programs. 
\end{acknowledgments}\

\end{document}